\documentclass[10pt,aps,prl,twocolumn,superscriptaddress,print]{revtex4-1}

\usepackage{lmodern}
\usepackage[version=3]{mhchem}
\usepackage{graphicx}
\usepackage{hyperref}
\usepackage{hypcap}

\hypersetup{ 
colorlinks=true,
linkcolor=blue,
citecolor=blue
}

\begin{document}

\title{Tunability of Room Temperature Ferromagnetism in Spintronic Semiconductors through Non-magnetic Atoms}

\author{Brett Leedahl}
\address{Department of Physics and Engineering Physics, University of Saskatchewan, 116 Science Place, Saskatoon, Saskatchewan S7N 5E2, Canada}

\author{Zahra Abooalizadeh}
\address{Department of Physics and Engineering Physics, University of Saskatchewan, 116 Science Place, Saskatoon, Saskatchewan S7N 5E2, Canada}

\author{Kyle LeBlanc}
\address{Department of Physics and Engineering Physics, University of Saskatchewan, 116 Science Place, Saskatoon, Saskatchewan S7N 5E2, Canada}

\author{Alexander Moewes}
\address{Department of Physics and Engineering Physics, University of Saskatchewan, 116 Science Place, Saskatoon, Saskatchewan S7N 5E2, Canada}
 

\begin{abstract}
The implementation and control of room temperature ferromagnetism (RTFM) by adding magnetic atoms to a semiconductor's lattice has been one of the most important problems in solid state state physics in the last decade. Herein we report for the first time, to our knowledge, on the mechanism that allows RTFM to be tuned by the inclusion of \emph{non-magnetic} aluminum in nickel ferrite. This material, NiFe$_{2-x}$Al$_x$O$_4$ (x=0, 0.5, 1.5), has already shown much promise for magnetic semiconductor technologies, and we are able to add to its versatility technological viability with our results. The site occupancies and valencies of Fe atoms (Fe$^{3+}$ T$_d$, Fe$^{2+}$ O$_h$,  and Fe$^{3+}$ O$_h$) can be methodically controlled by including aluminum. Using the fact that aluminum strongly prefers a 3+ octahedral environment, we can selectively fill iron sites with aluminum atoms, and hence specifically tune the magnetic contributions for each of the iron sites, and therefore the bulk material as well. Interestingly, the influence of the aluminum is weak on the electronic structure (supplemental material), allowing one to retain the desirable electronic properties while achieving desirable magnetic properties.

\end{abstract}

\maketitle 

\section{Introduction}
Spinel oxides (AB$_2$O$_4$) often have quite unique and highly tunable and versatile functionalities.\cite{Garlea2008,Liu2005} Among spinel oxides, ferrites are emerging as a viable magnetic material for use in novel technologies; especially in the area of spintronics, wherein magnetic semiconductors play a central role in generating highly spin-polarized currents.\cite{Kajiwara2010, Li2016} Indeed, NiFe$_2$O$_4$ films have been shown to display spin-polarized currents, and adjustable electrical properties through varying growth conditions.\cite{Luders2006,Liu2013,Jing2011} Currently, nickel ferrites are extensively used in a number of electronic devices because of their high magnetic permeabilities, high electrical resistivity, mechanical hardness, chemical stability, and reasonable cost.\cite{Haneda1987} Understanding the role of electron correlation effects in these ferrites has been a major challenge.

Theoretical studies have suggested that NiFe$_2$O$_4$ has Ni ions exclusively on $B$ octahedral (O$_h$ in point group representation) sites, and Fe ions distributed equally among $A$ tetrahedral (T$_d$ in point group representation) and $B$ sites (referring to the AB$_2$O$_4$ notation).\cite{Cullity2009} On one hand, a strength of NiFe$_2$O$_4$ is that its properties can be tuned based on synthesis conditions, but on the other, measurements of its properties have shown a variety of results. For example, it has been reported to have a magnetic moment in ultra-thin films that is 2.5 times larger than in the bulk.\cite{Luders2005} Multiple studies have investigated the properties of NiFe$_2$O$_4$, but the reported observations lack consistency. \cite{Sun2012, Dolia2006, Haetge2010,Balaji2005,Szotek2006} These discrepancies make it a worthwhile endeavour to pursue complementary techniques (x-ray, as opposed to optical or theoretical methods) to add to the body of work for such a technologically important material.

The electronic and magnetic effects of alloying different elements (such as Al ions) into nickel ferrite is a topic that warrants further exploration. While the magnetism due to Ni atoms in NiFe$_2$O$_4$ was thoroughly studied,\cite{vanderlaan1999} such non-magnetic alloying provides a promising pathway to tuning its magnetic properties, which is highly desired in the field of spintronics.\cite{Green2015,Wang2015}

Previously, the effect of Al substitution on NiFe$_2$O$_4$ was shown to cause both the Curie temperature ($T_C$) and lattice constant to decrease slightly with increasing Al concentration.\cite{Mozaffari2011} In the present study the effect that Al doping has on nickel ferrite alloys is explored by using soft x-ray spectroscopy techniques. The x-ray absorption spectroscopy (XAS) at the $L_{2,3}$-edges of Fe and Ni allowed us to examine their element specific electronic and magnetic structures.\cite{Boyko2013,Perez2014} Finally, through comparison of the experimental spectra and crystal field multiplet calculations of transition metal $L_{2,3}$-edges spectra, we were able to extract the local coordination of these atoms.

\section{Experiment and Calculation Details}
Nanocrystalline powders of NiFe$_{2-x}$Al$_x$O$_4$ (x=0.0, 0.5, 1.5) were prepared by the sol-gel method; detailed information regarding the synthesis of these materials can be found in a previous publication \cite{Mozaffari2011}. After deposition, the powders were annealed separately in air at different temperatures from 400$^{\circ}$C to 1100$^{\circ}$C for two hours in order to get the final single phase products. Lastly, x-ray diffraction (XRD) scans were performed to ensure the single phase structure.

The crystal field multiplet calculations in this work use the algorithm initially formulated by Cowan, and the working code subsequently expanded on by Haverkort and Green \emph{et al}.\cite{Cowan1968, Green2011, Haverkort2014, Lu2014} The free parameters include the crystal field strength (from which the local symmetry can be deduced), oxidation state, and the scaling of the intra-atomic Coulomb and exchange (Slater) integrals. The dipole transition matrix elements calculated by this code are then used in the Kramers-Heisenberg equation to simulate spectra.\cite{Sakurai1967} All spectra are broadened by convolutions with a Lorentzian function (to simulate lifetime broadening), and a Gaussian function (to simulate experimental broadening) to match experimental conditions.

\begin{figure}
\begin{center}
\includegraphics[width=3.375in]{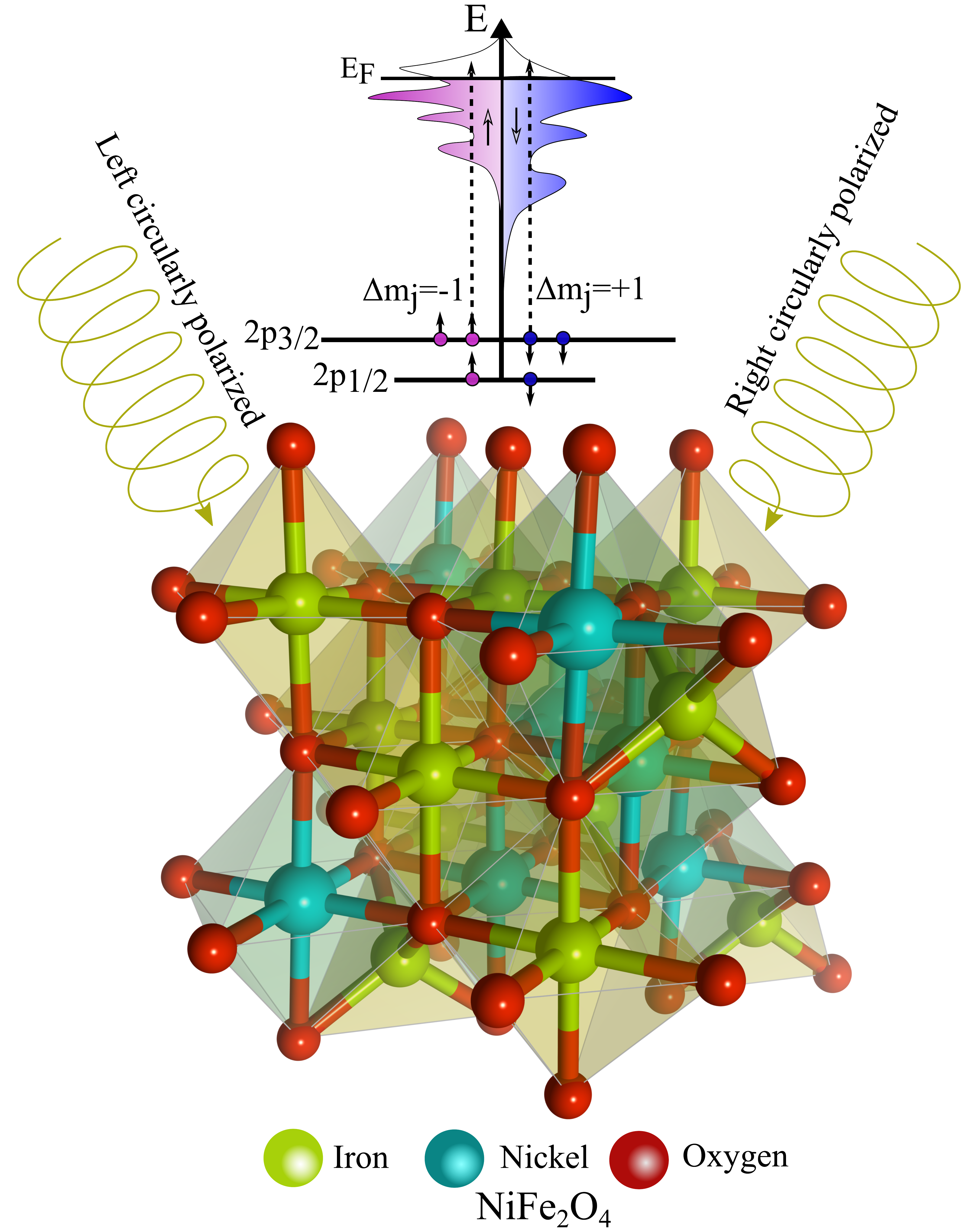}
\caption{XMCD is an element- and orbital-specific technique. It relies upon left and right circularly polarized x-ray to probe \emph{exclusively} the Fe $2p$ electrons. Under a magnetic field the spin up and spin down $2p$ electrons are disproportionately excited into the partially $3d$ band due to the difference in the unoccupied spin up and spin down states. Left and right polarized photons transfer $-\hbar$ and $\hbar$ angular momentum, respectively, to the excited electrons. Dipole selection rules govern the proportion with which spin up and spin down electrons are excited. Consequently, this information can be gathered in relatively simple XMCD sum rules,\cite{Carra1993} and the difference between left and right polarized absorption spectra determines the orbital and spin magnetic moments per Fe atom.\cite{Thole1992}}
\label{fig:xmcdscheme}
\end{center}
\end{figure}

Our x-ray magnetic circular dichroism (XMCD) measurements were performed at the REIXS Beamline of the Canadian Light Source mounted on a 0.5 T magnet to saturate the magnetic moments of the sample such that XMCD selection rules are valid. The x-ray photons used were incident at 45$^{\circ}$  to the sample normal, and maintained greater than 95\% circular polarization. A schematic of the XMCD process is shown in Fig. \ref{fig:xmcdscheme}. Using this technique we were able to decompose the magnetic signal of our samples into different symmetries and oxidation states of iron.

\section{Results and Discussion}
\begin{figure*}
\begin{center}
\includegraphics[width=6.5in]{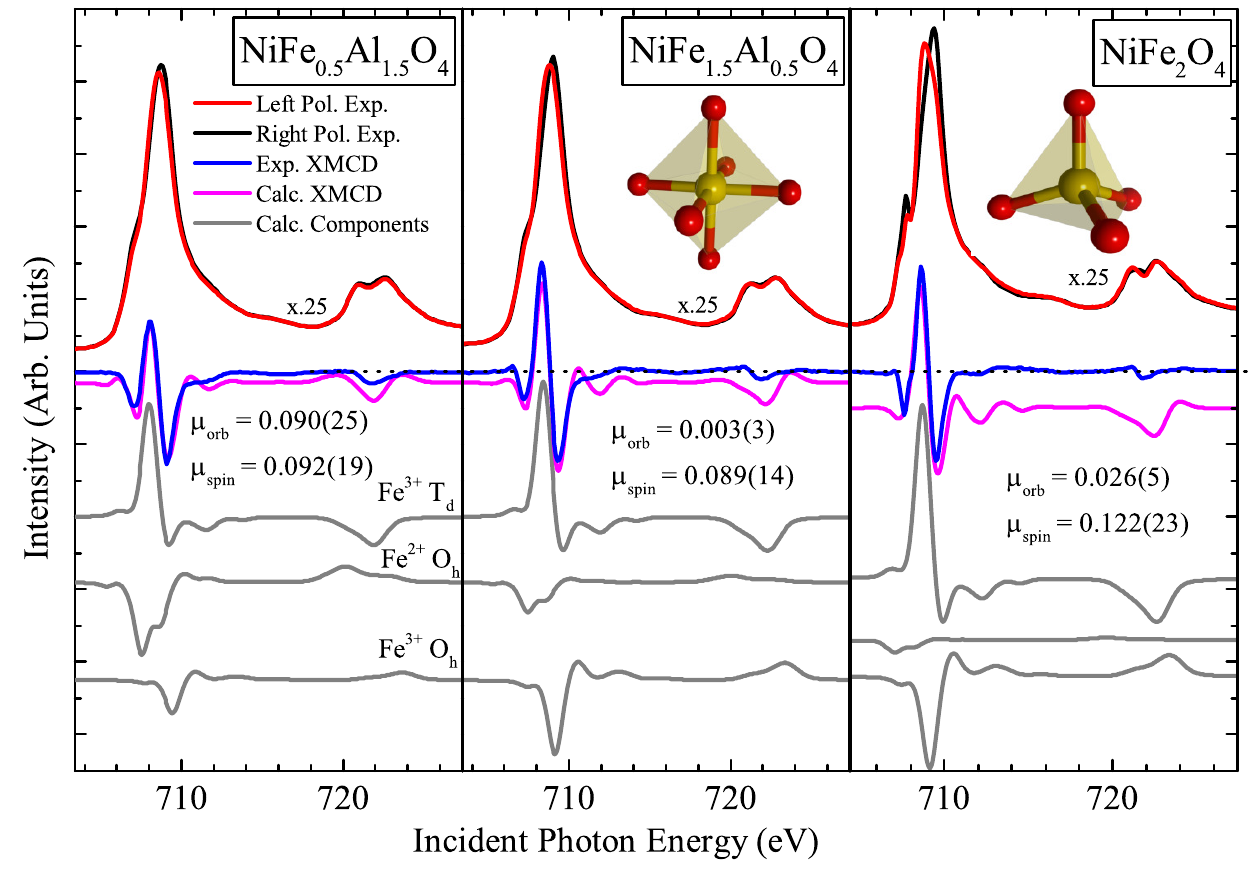}
\caption{Calculated (magenta) and experimental (blue) XMCD spectrum at the Fe $L_{2,3}$-edge with both left and right circularly polarized x-rays. Grey curves (from top to bottom) show calculated Fe$^{3+}$ T$_d$, Fe $^{2+}$ O$_h$,  and Fe$^{3+}$ O$_h$ components of the XMCD spectra, respectively. With increasing Al content we see the following trends: mildly decreasing Fe$^{3+}$ T$_d$ signal, largely increasing Fe$^{2+}$ O$_h$ signal, and a steadily decreasing Fe$^{3+}$ O$_h$ signal. The experimentally derived spin and orbital magnetic moments for each sample are shown in their respective panels in units of bohr magnetons. The experimental error is also shown in bracket notation; note that this does not account for approximations made within the sum rules themselves.}
\label{fig:xmcd}
\end{center}
\end{figure*}

The measured XMCD at the Fe $L_{2,3}$-edges for all samples are shown in Fig. \ref{fig:xmcd}, taken in total electron yield (TEY) mode. The left and right (red and black) polarized XAS spectra are scaled by a factor of 0.25 compared to the XMCD signal (blue) for clarity; this XMCD signal is the difference between the two absorption spectra. In addition to the experimental spectra, we have included crystal field multiplet calculated spectra for the three individual components (Fe$^{3+}$ T$_d$, Fe $^{2+}$ O$_h$,  and Fe$^{3+}$ O$_h$), along with an overall calculated XMCD spectrum, computed as a linear sum of these components.\cite{Pattrick2002} It is also worthy to note that the absorption spectra were also measured in bulk sensitive inverse partial fluorescence yield mode, and hence free of saturation and self-absorption effects that are known to alter feature intensities when using other XAS techniques like fluorescence detection.\cite{Achkar2011} These spectra agreed with our TEY spectra in Fig. \ref{fig:xmcd}, however they are inherently substantially noisier and the XMCD sum rules are not possible to use reliably with noisy data.
  
The Fe XMCD spectra comprises of a superposition of the three main components that are derived from the three sites occupied by iron: Fe$^{2+}$ octahedral ($d^6$O$_h$), Fe$^{3+}$ tetrahedral ($d^5$T$_d$), and Fe$^{3+}$ octahedral ($d^5$O$_h$). The Fe$^{3+}$ ions at the tetrahedral sites are coupled antiferromagnetically to those at the octahedral sites. This antiferromagnetic coupling is clear because in order to achieve agreement with experiment, a sign reversal of the spin operators was required in the calculations.\cite{Pattrick2002}
  
What we discovered was that an exciting trend emerges among the intensities of the three components. As the Al content increases we observe that both Fe$^{3+}$ signals decrease in magnitude, while the Fe$^{2+}$ XMCD signal increases. This is in accordance with what we would expect from the argument that Al strongly prefers to be in a 3+ oxidation state, and so tends to replace Fe$^{3+}$ atoms. Furthermore, as observed previously,\cite{Mozaffari2011} we also concur that the Al$^{3+}$ atoms tend to prefer the octahedral environment of Fe, and therefore the Fe O$_h$ signal is considerably suppressed compared to that of the tetrahedral sites, which are only mildly diminished. As a result of these two strong preferences (Al into O$_h$ and 3+ sites), the Fe$^{3+}$ O$_h$ signal largely dies out, while the Fe$^{2+}$ ions become a large contributor to the magnetism with increasing Al content.

That is to say, as Fe$^{3+}$ O$_h$ sites become filled with Al$^{3+}$ ions, this site's contribution to the overall ferromagnetism is gradually reduced until it is nearly zero. In a similar, but less drastic way, Fe$^{3+}$ T$_d$ sites are filled by Al$^{3+}$ ions. Surprisingly this does not necessarily imply that the overall magnetism of the material must be reduced accordingly. Only by viewing the XMCD signals in Figure \ref{fig:xmcd} can we explain this phenomena. The fingerprint XMCD signals of the two reduced Fe$^{3+}$ sites largely (but by no means completely) cancel one another out. Consequently, their simultaneous reduction does not manifest itself so drastically in the material's bulk magnetic properties. As a matter of fact, what we found is that Fe$^{2+}$ O$_h$ sites emerge as a significant contributor the overall ferromagnetism when the other two sites are reduced. It is this interplay between the three Fe sites and their relative occupancies---which we can \emph{only} discern via XMCD---that gives rise to the bulk magnetic properties. This finding is validation of the power of XMCD, as well as illustrating an important technique that could be adopted and applied to reach its full potential in the realm of synthesizing spintronic devices, wherein the tuning of magnetic moments is of the utmost importance.

From our experimental XMCD spectra we determined the magnetic moments of the samples using the left and right circularly polarized XAS and XMCD sum rules (orbital and spin moments are shown in Fig. \ref{fig:xmcd}, with errors due to experiment shown in bracket notation).\cite{vanderLaan2013} By adding the orbital $\mu_{orb}$ and spin  $\mu_{spin}$ moment of each sample ($\mu$ = 0.148$\mu_B$ for x = 0, $\mu$ = 0.092$\mu_B$ for x = 0.5, and $\mu$ = 0.182$\mu_B$ for x = 1.5), one can see that the net magnetization decreases when the Al content is increased to x = 0.5, and then increases for x = 1.5. This is consistent with our conclusion that the interplay between the three XMCD signals is what gives rise to the observed bulk magnetic properties. The reduction of some magnetic sites in turn may give rise to the appearance of others, leading to a complicated exchange between them, and not just a simple reduction in magnetism as magnetic atoms are replaced by non-magnetic atoms. 

As an additional point it should be noted that there are limitations to using the sum rules in determining precise quantitative values of spin and orbital magnetic moments. (1) Experimental errors such as noise, and the fact that left and right polarized XMCD spectra cannot be taken simultaneously lead to uncertainty in the integrated XMCD spectra (for which small experimental errors can propagate into relatively large absolute quantitative errors, these are the errors shown in Fig. \ref{fig:xmcd} brackets). (2) Approximations made within the sum rules themselves such as: assuming the spin-quadrupole coupling term is zero (which is commonly used for transition metal $L$-edges),\cite{Piamonteze2009} and the uncertainty in the number of $d$-electron holes, which will vary due to some degree of covalency and mixing of oxidation states.\cite{Wasinger2003} For these reasons, the absolute values of our magnetic moments are of secondary importance. What is important for the proper analysis of our data is identifying and explaining the trends and contributions of each of the Fe sites as the amount of Al varies.


Hence, we have found that small changes in the site occupancies can give rise to considerable differences in the relative peak intensities of the XMCD. Using the three calculated components, it is possible to predict the spectral shapes of spinels with different ratios of Fe$^{2+}$/Fe$^{3+}$ at the two sites (octahedral/tetrahedral O$_h$/T$_d$). This principle can be understood better by considering the \emph{inversion parameter, i}---it tells us the fraction of Fe ions in T$_d$ and O$_h$ sites, and always takes a value between zero and one. It can be written in the following form:
\[[\textup{Ni}_{1-i}\textup{Fe}_{i}]^{\textup{T}_d}[\textup{Ni}_i\textup{Fe}_{2-i}]^{\textup{O}_h}\textup{O}_4\]
Therefore, a material with an inversion parameter of 1.0 would contain Fe$^{3+}$ in both octahedral and tetrahedral sites in a 1:1 ratio, while a normal spinel ($i = 0$) ferrite of the same formula would contain Fe$^{3+}$ in octahedral sites only. Indeed, in the right panel of Fig. \ref{fig:xmcd} we see that the Fe$^{2+}$ is quite negligible, and so NiFe$_2$O$_4$ has an inversion parameter very close to one, as has been previously found using other methods.\cite{Carta2009}

A superposition of the three theoretical components can therefore be fit to the experimental spectra, producing site occupancy ratios of Fe at the three sites. Note that Fe$^{2+}$ at T$_d$ sites have been ignored---if it is included in the fitting process, a small component of $<$ 0.1 atoms per unit formula may be present, but is not significant.\cite{Richter2009} The small discrepancies between experiment and calculation can be attributed to the long range effects of the crystal field due non-nearest neighbours and the addition of Al atoms to the host lattice, as well as slight distortions from spectra being taken in TEY mode.

Thus, we can deduce our principal revelation from a few basic tenets. (1) Iron is frequently found in many magnetic compounds in some combination of its four most common environments (Fe$^{2+}$ octahedral, Fe$^{2+}$ tetrahedral, Fe$^{3+}$ octahedral, and Fe$^{3+}$  tetrahedral. (2) Each of these four sites has a unique magnetic signature that can be measured via XMCD. (3) We can then exploit the fact that many elements strongly tend to a given oxidation state and local symmetry. For example, aluminum atoms are found nearly exclusively in a 3+ oxidation state and in octahedral environments. Hence, upon addition of these Al atoms to some host lattice, they will preferentially replace atoms in 3+ octahedral sites, and to a lesser degree 3+ tetrahedral sites. The key point is that the Al atoms \emph{will not} substitute into 2+ sites. Herein, we have shown that it is feasible to exploit this property with the replacement of magnetic Fe atoms by non-magnetic Al atoms in NiFe$_2$O$_4$.  Therefore, by adding aluminum (or other non-magnetic atoms) it is possible to tune the site occupancy ratios of the ferromagnetic atoms, leading to a tuning of the magnetism of the compound as a whole. What is even more amazing about the fine tuning of the magnetism, is that it is all accomplished while retaining the host material's electronic properties; a full discussion and analysis of the electronic properties is given in the \emph{Supplemental Material}.\cite{supplemental}

\section{Conclusion}
With room temperature ferromagnetic materials becoming a burgeoning area of research in recent years, it is required that substantial advances in the control and understanding of magnetic properties are achieved.\cite{Kodama1996} The lesser studied idea of using non-magnetic atoms offers a novel avenue of departure from the more customary iron/nickel doping. We have shown that ferromagnetic single-phase nickel ferrite NiFe$_{2}$O$_4$ can have the occupancies of the three iron environments adjusted by the inclusion of aluminum atoms, hence altering the spin and orbital magnetic moments of the bulk material. This was shown to be possible only through the use of synchrotron-based XMCD spectroscopy, alongside crystal field multiplet calculations. Our study shows a proof of concept that by decomposing the ferromagnetism into its constituents, we can make substantial advances in understanding the source of the magnetism. In turn this will surely lead to corresponding advances in the tailoring of magnetism that can be achieved with careful synthesis. This ought to garner further interest in the popular field of spintronic devices, wherein controlling electron spin has been one of the most important topics in condensed matter physics in recent decades.

\section{Acknowledgements}
This work was supported by the Natural Sciences and Engineering Research Council of Canada (NSERC) and the Canada Research Chairs program. Measurements were performed at the Canadian Light Source (supported by NSERC, the Canadian Institutes of Health Research, the Province of Saskatchewan, Western Economic Diversification Canada, and the University of Saskatchewan) and the Advanced Light Source (supported by the U.S. Department of Energy under Contract No. DE-AC02-05CH11231). We are also grateful to Dr. M. Mozaffari and Prof. J. Amighian for their cooperation in the sample synthesis process.




%

\end{document}